\documentclass[preprint]{aastex}

\newcommand{\etal}{et~al.}
\newcommand{\kms}{km~s$^{-1}$}

\newcommand{\CIVdblt}{\ion{C}{4}~$\lambda\lambda 1548, 1550$}
\newcommand{\NVdblt}{\ion{N}{5}~$\lambda\lambda 1393, 1403$}
\newcommand{\OVIdblt}{\ion{O}{6}~$\lambda\lambda 1032, 1038$}
\newcommand{\SiIVdblt}{\ion{Si}{4}~$\lambda\lambda 1394, 1403$}
\newcommand{\AlII}{\ion{Al}{2}}
\newcommand{\CII}{\ion{C}{2}}
\newcommand{\CIII}{\ion{C}{3}}
\newcommand{\CIV}{\ion{C}{4}}
\newcommand{\HeII}{\ion{He}{2}}

\newcommand{\SiIV}{\ion{Si}{4}}
\newcommand{\SiII}{\ion{Si}{2}}
\newcommand{\NV}{\ion{N}{5}}

\newcommand{\OVII}{\ion{O}{7}}
\newcommand{\OVIII}{\ion{O}{8}}

\newcommand{\Lya}{{Ly}\,$\alpha$}
\tighten
\begin{document}

\singlespace


\slugcomment{{\it The Astrophysical Journal Letters}}
\shortauthors{~~Ganguly \etal}
\shorttitle{MRC~2251--178}

\title{\bf Variable UV Absorption in the Spectrum of MRC~2251--178
\footnote{Based on observations with the NASA/ESA {\it Hubble Space Telescope},
obtained at the Space Telescope Science Institute, which is operated by AURA,
Inc., under NASA contract NAS 5--26555.}}

\pagestyle{empty}

\author{Rajib~Ganguly, Jane~C.~Charlton, Michael~Eracleous}
\medskip
\affil{\normalsize\rm Department of Astronomy and Astrophysics \\
       The Pennsylvania State University, University Park, PA 16802 \\
       e-mail: {\tt ganguly, charlton, mce@astro.psu.edu}}
\pagestyle{empty}
\begin{abstract}
We present an ultraviolet spectrum of {MRC~2251--178} taken with
HST/STIS-G230L. The observation is part of a snap shot program of QSOs
to systematically search for intrinsic absorption lines through
variability. The sample consists of all QSOs observed with HST/FOS
which showed associated narrow absorption lines. The FOS spectrum of
{MRC~2251--178}, taken in 1996, showed an associated {\CIV} doublet
with an {$\lambda1548$} equivalent width {$1.09\pm0.09$~\AA}
{\citep{mon01}}. It is not detected in the STIS spectrum taken four
years later, down to a {$3\sigma$} threshold of {$0.19$~\AA}. In
addition to its other accolades, these observations make
{MRC~2251--178} the very first QSO at low redshift in which the
associated absorption is shown to be truly intrinsic. We discuss the
implications of this and suggest courses for future study.
\end{abstract}


\section{Introduction}
\pagestyle{myheadings}

The identification of QSO-intrinsic narrow absorption
lines\footnote{Hereafter, we abbreviate narrow absorption line with
the acronym NAL and refer to a QSO that hosts a truly {\it intrinsic}
NAL as a NALQSO.} has exploded over the past four years with the
advent of high resolution spectroscopy with large ground-based
telescopes. With this explosion has come new insights on how to
identify these absorbing systems - that is, how to separate them from
randomly distributed intervening material. Historically, large
ensembles of intrinsic absorbers have been identified using
statistical arguments, either demonstrating an excess of absorbers in
some redshift path {\citep{foltz86,and87}} or by correlating the
velocity distribution (with respect to the QSO emission redshift) of
absorbers with QSO properties {\citep{rich99}}. Such surveys have
demonstrated that, in high redshift QSOs, there {\it is} an excess of
absorbers that lie within {5000~\kms} of the QSO emission redshift
(termed ``associated'' absorbers). It is now generally accepted that a
large fraction of associated absorbers are indeed intrinsic. However,
it is not truly known how large this fraction is. Moreover,
\citet{rich99} demonstrated that a significant fraction of
non-associated (that is, high ejection velocity) absorbers may also be
intrinsic.

The two smoking guns, according to \citet{bs97}, that identify an
intrinsic system are: (1) variability of the spectral profiles and/or
equivalent widths over time [e.g.,
\citet{ham95,bhs97,ham97b,ham97c,ald97}]; and (2) the signature of
partial coverage [e.g., \citet{bs97,bhs97,ham97a,gan99}].  Thusfar,
all cases of confirmed QSO-intrinsic NALs are at high redshift, where
the resonant UV doublets of metals are shifted into the optical. At
low redshift, \citet{gan01} estimate from statistical excesses that
{$\sim25$\%} of QSOs have intrinsic NALs. As a first step in
determining that true fraction of intrinsic NALs, and deciphering the
dominant cause of variability (if one exists) we have undertaken an
HST/STIS snap shot program to identify intrinsic NALs in a sample of
QSOs where associated NALs have been detected in archived HST/FOS
spectra. In this effort, we also hope to look for high ejection
velocity absorbers at low redshift and decipher the nature of the
variability. We note that, up until now, only six QSOs have been shown
to host intrinsic absorption through time variability - all at
{$z_{\mathrm{em}}>1.5$}. Furthermore all cases of associated, time
variable absorption have been been low luminosity AGN
{\citep{wal90,ss93,mar96,wsgh97,cren00}}.

In this letter, we report the first confirmation at low redshift of an
associated NAL that is truly intrinsic to a QSO, {MRC~2251--178}
[$z_{\mathrm{em}}= 0.066092$; \citet{bbdt83}]. The
radio-quiet {MRC~2251--178} is well-known for having a highly
variable X-ray warm absorber, the first one known
{\citep{hal84}}. \citet{mon01} first reported an associated UV
absorber and, assuming that the same gas produced the X-ray warm
absorption, inferred that the mass loss rate from the disk and the
accretion rate were comparable.

In \S2, we present our HST/STIS snapshot of {MRC~2251--178}. In \S3,
we show that the associated UV absorption has varied over the past
four years (thus, verifying its intrinsic nature). In \S4, we discuss
the implications of variability on analyses of intrinsic absorption
lines and make suggestions for future work on {MRC~2251--178}

\section{Data}

The sample for our snap shot program is drawn from QSOs in the HST/FOS
archive. The sample is restricted to only those QSOs which have been
observed with one of the ``high'' resolution ($\sim230$~\kms) FOS
gratings (G130H, G190H, G270H) and for which there is evidence of
associated absorption in the form of a {\CIVdblt}, {\NVdblt},
{\OVIdblt}, or {\Lya} feature. The sample, comprised of 37 QSOs,
includes 13 NALQSO candidates from {\citet{gan01}}, which derived from
a subsample of the {\it HST Quasar Absorption Line Key Project QSOs},
as well as 24 QSOs which were not part of the Key Project. The QSOs
from the Key Project {\citep{kpi,kpii,kpvii,kpxiii}} were obtained
from D. Schneider, B. Jannuzi, and S. Kirhakos. The non-Key Project
QSOs were reduced and supplied by S. Kirhakos. All QSOs were obtained
fully reduced with effective continuum\footnote{The effective
continuum is the sum of the continuum and broad emission lines from
the QSO. It is the spectrum that is presumably incident on the
absorbing gas.} fits according to {\citet{kpii}}.

\begin{figure}
\figurenum{1}
\plotone{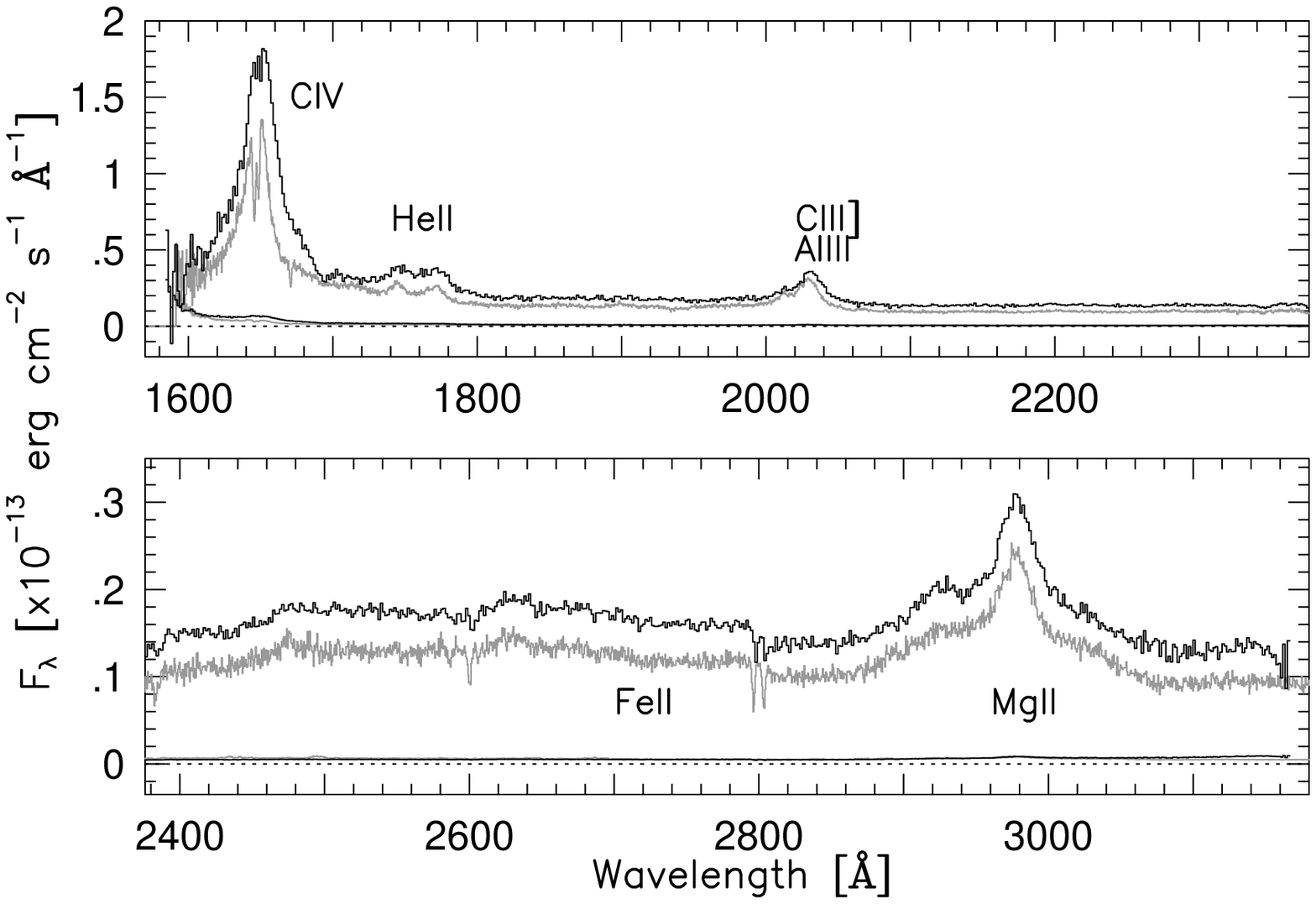}
\figcaption[fig1.eps]{STIS-G230L spectrum of {MRC~2251--178}. The two
histograms shown are the STIS spectrum (dark) from 5 November 2000 and
FOS spectrum (light) from 2 August 1996 \citep{mon01}. The STIS
spectrum covers the range {1570--3180~\AA}. The emission lines covered
include {C~{\sc iv}}, {He~{\sc ii}}, {C~{\sc iii}]}, {Al~{\sc iii}},
the {Fe~{\sc ii}} complex, and {Mg~{\sc ii}}. Galactic absorption from
{Fe~{\sc ii}} and {Mg~{\sc ii}} as well as a hint of the associated
{C~{\sc iv}} absorber are also present.}
\label{fig:fullspec}
\end{figure}

The primary goal of the snap shot program is to detect variability of
associated NALs. To that end, the QSOs are observed using HST/STIS
with the G230L grating and the {$52\times0.2$''} slit. The dispersion
of the grating is {1.58~\AA} per pixel, with two pixels per resolution
element. This corresponds to {300--600~\kms} across the spectrum,
compared to the roughly constant {230~\kms} resolution of the FOS
spectra. In Fig~\ref{fig:fullspec}, we show the full spectrum of
{MRC~2251--178} (darker histogram) resulting from a {500~s} exposure
taken on 5 November 2000. The signal-to-noise per pixel at {2376~\AA}
(the central wavelength) is {$\sim25$}. The comparison HST/FOS
spectrum, also shown in Fig~\ref{fig:fullspec} (lighter histogram),
was taken on 2 August 1996 with all three {230~\kms} resolution
gratings {\citep{mon01}}. The relevant portion of the spectrum (taken
with the G190H grating) has a signal-to-noise similar to the STIS
spectrum. The flux calibration from the reduction pipelines for both
the STIS-G230L and FOS-G190H spectra are accurate to within 3\%.  The
wavelength calibration for the FOS-G190H spectrum is accurate to 0.11
diodes (0.2~\AA) while that of the STIS-G230L spectrum is accurate to
0.2--0.5 pixels (0.3--0.8~\AA). A STIS-G140M spectrum, obtained 3
February 1998 by J. Stocke, spans the wavelength range
{1194--1300~\AA} (in two separate grating tilts). Although this
spectrum covers the associated absorption from {\Lya} and {\NV}
transitions, there is, unfortunately, no overlap with our
STIS-G230L. As a result, no direct comparison can be made to the
STIS-G140M spectrm.

\section{Analysis \& Results}

It is apparent by visual inspection of the STIS
(Fig.~\ref{fig:fullspec}) and FOS (Monier {\etal}~2001: Fig. 1)
spectra that there has been a change in the state of both the emission
and associated absorption lines of {MRC~2251--178}. The STIS spectrum
has a generally higher flux and the associated {\CIV} absorption has
apparently disappeared. To examine the magnitude of the variability,
we convolved the FOS spectrum with a Gaussian ({575~\kms} FWHM) and
resampled the result with a {1.58~\AA/pixel} dispersion to mimic the
expected STIS-G230L spectrum around the {\CIV} emission line.

\begin{figure}
\figurenum{2}
\plotone{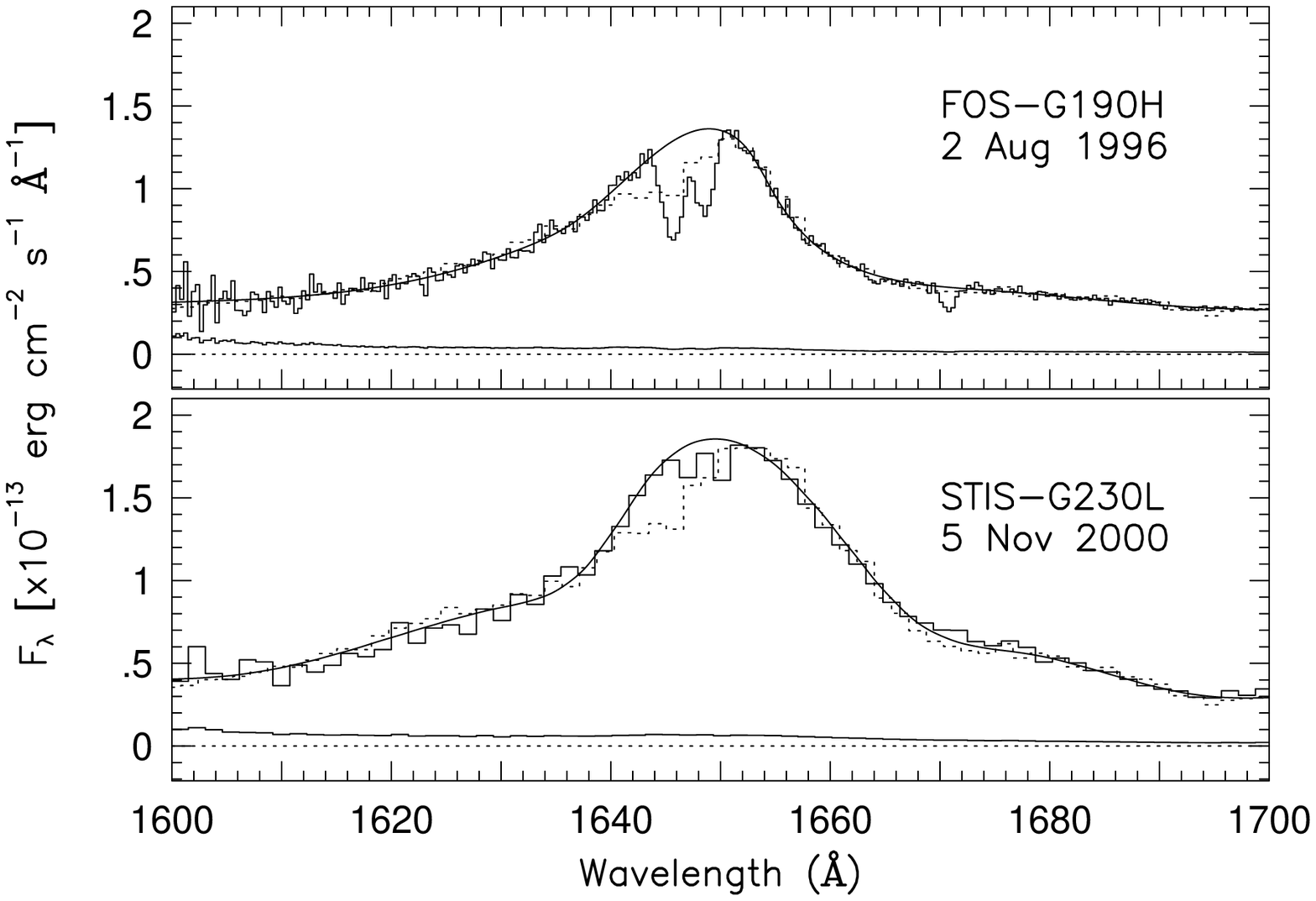}
\figcaption[fig2.eps]{Comparison of {FOS-G190H} and {STIS-G230L} spectra. In
both panels we show the flux density and error spectrum in the region
around the {C~{\sc iv}} emission line. In the top panel we show the
FOS-G190H spectrum (solid histogram), the fit to the effective
continuum (smooth solid spline), and the expected STIS-G230L spectrum
if nothing had varied (dotted histogram). In the bottom panel, we show
the observed STIS-G230L spectrum (solid histogram), the fit to the
effective continuum (smooth solid spline), and the expected spectrum
if no absorber properties changed (dotted histogram). It is clear,
to about {$10\sigma$} confidence, that the associated {C~{\sc iv}}
absorber varied.}
\label{fig:compare}
\end{figure}

In Fig.~\ref{fig:compare}, we show the {\CIV} emission and associated
absorption lines from the observed FOS spectrum (top panel; solid
histogram), and the observed STIS spectrum (bottom panel; solid
histogram). Overplotted on each are the effective continuum fits
(smooth curve) and the expected STIS spectrum (dotted histograms). In
the top panel, the expected STIS spectrum reflects what should have
been observed if there were no variability in either the {\CIV}
emission or associated absorption lines.  In the bottom panel, the
expected STIS spectrum shows what should have been observed if there
were no change in the properties of the associated absorber (that is,
if only the shape of the {\CIV} emission line and the continuum flux
changed).  With STIS-G230L, the associated {\CIV} absorption is
sampled by seven pixels (3.5 resolution elements) and should have been
detected in the observed spectrum. (The Galactic {\AlII} line at
{1670~\AA}, which is sampled by less than a pixel in the STIS
spectrum, is completely washed out by the instrumental profile, making
it indistinguishable from the {\CIV} and {\HeII} emission lines.)
Furthermore, {\citet{mon01}} measured a {\CIV$\lambda1548$} equivalent
width of {$1.09\pm0.09$~\AA} in the FOS spectrum. Using the unresolved
feature detection method from {\citet{kpii}}, the absorber is not
formally detected in the STIS spectrum down to a {$3\sigma$}
equivalent width limit of {$0.19$~\AA}. This is more than a
{$10\sigma$} change. It is, therefore, clear and robust that there has
been variability in the properties of the associated absorber.

\section{Discussion}

\subsection{Variability in the UV spectrum}

The change in the state of associated {\CIV} absorption happened
sometime in the last four years. Although, we cannot say accurately on
what timescale that variability occurred, it is now safe to say that
this is the first clear demonstration of truly intrinsic narrow
absorption in a low redshift QSO. Variability in absorption lines is
generally attributed to one of two scenarios. In the first scenario,
bulk motion, the absorption gas has an appreciable motion transverse
to the line of sight and the total column density through the gas
changes with time. The observational signature is that the equivalent
width of all lines due to this gas should change in the same way (that
is, all get stronger or all get weaker) with time.  In the second
scenario, ionization/recombination, the physical state of the absorber
changes such that the absorbing gas becomes more or less ionized with
time.  Observationally, this would mean that the equivalent widths of
higher-ionization lines and lower-ionization lines should change in
opposite directions. This assumes, however, that the kinematics of the
gas does not change so that the column densities of the ions drive the
changes in the equivalent widths. Unfortunately, the four year
separation of the two UV observations and the lack of coverage of
relevant species by the STIS G230L observation (e.g., {\Lya}, \NVdblt,
\SiIVdblt, \SiII$\lambda1260$, \CII$\lambda1335$) do not allow one to
distinguish between these two causes of the variability in the
spectrum of {MRC~2251--178}.

Nevertheless, since we know that the variability timescale is less
than 4~years, we can provide an upper limit on the distance between
the absorbing material and the ionizing sources. If we assume that the
cause of variability is recombination of {\CIV} to {\CIII} and that
the temperature of the absorber is about {$T_{\mathrm{e}} = 2 \times
10^5$~K} as in UM 675 {\citep{ham95}}, then the minimum density is
{$n_{\mathrm{e}} = 1/t_{\mathrm{rec}} \alpha >
3000~\mathrm{cm}^{-3}$}, where the recombination coefficient is
{$\alpha = 2.8 \times 10^{-12}~\mathrm{cm}^3~\mathrm{s}^{-1}$}
{\citep{ar85}}. This implies a maximum distance of 2.4~kpc, if {\CIV}
is the dominant species. This is consistent with the estimate from
{\citet{mon01}}. While not very restrictive, the distance estimate
does rule a number of candidates for the absorbers, including the
intergalactic medium of the host cluster and most of the interstellar
medium of the host galaxy. The remaining possibilities are the material
mixed with broad emission line region (BELR), material outside the BLR,
and outflowing/ejected material.

\subsection{Suggestions for Future Work}

The next step in understanding the intrinsic absorber in
{MRC~2251--178} is to decipher the cause of variability - bulk motion
or ionization/recombination - and then determine if there is a direct
physical relationship between the gas and the X-ray warm absorber.  In
principle, these issues can be tackled, but they require a regimented
observing schedule. To efficiently determine both the underlying cause
of variability and the geometry of absorbing gas relative to the
central engine, periodic observations at medium to high resolution
should be obtained of the rest-frame wavelength range {1200--1600~\AA}
(e.g., with the E140M echelle on STIS). This covers transitions of
\SiII, \Lya, \NV, \CII, \SiIV, and \CIV. The frequency of the
observations should be less than (or on the order of) once in two
months. If the QSO and/or absorber properties vary within the course
of observations, this schedule should provide the variability
timescale.  In addition, at high resolution, photoionization modeling
of the individual components should constrain the physical conditions
of the gas. Furthermore, measurements of the partial coverage fraction
are possible (and necessary) with the resolved components. The
combination of the partial coverage fraction, the physical conditions
of the gas (e.g. the ionization parameter, density), and the
variability timescale can constrain the distance from the central
engine, the size, and the transverse velocity of the absorbers.

Once the range of conditions in the UV absorbing gas are known, one
can then consider its relationship to the X-ray warm absorber. This,
however, is difficult since the X-ray warm absorber is known to vary
on short timescales ($<1$~year; Halpern 1984\nocite{hal84}). Thus,
simultaneous observations of {MRC~2251--178} in both the UV and the
X-ray are required so that inferred relationships are not affected by
the changing conditions. In the X-ray, observations can be carried
out with either the Chandra/HETG or XMM-Newton/RGS to cover {\OVII}
and {\OVIII}. Ultraviolet observations should be carried out with
either HST/STIS-E140M, which covers a range of ionization states,
and/or with FUSE, which covers the {\OVIdblt} doublet and the Lyman
series.

\acknowledgements

Support for this work was provided under grants HST--GO--08681.01--A
and STSI AR--08763.01--A. We are grateful to Sofia Kirhakos for
providing the uniformly and fully reduced FOS archive with continuum
and emission line fits.


\begin{thebibliography}{}

\bibitem[Aldcroft {\etal}(1997)]{ald97}
Aldcroft, T., Bechtold, J., \& Foltz, C. 1997, in ASP Conference
Ser. 128, Mass Ejection from Active Galactic Nuclei, ed. N. Arav,
I. Shlosman, \& R.  Weymann (San Francisco: ASP), 25

\bibitem[Anderson {\etal}(1987)]{and87}
Anderson, S. F., Weymann, R. J., Foltz, C. B., \& Chaffee, F. H. 1987, 
\aj, 94, 278

\bibitem[Arnaud \& Rothenglug(1985)]{ar85}
Arnaud, M., \& Rothenflug, R.
1985, \aaps, 60, 425

\bibitem[Barlow {\etal}(1997)]{bhs97}
Barlow, T. A., Hamann, F., \& Sargent, W. L. W., 1997, in ASP
Conference Ser. 128, Mass Ejection from Active Galactic Nuclei,
ed. N. Arav, I. Shlosman, \& R. Weymann (San Francisco: ASP), 13

\bibitem[Bahcall {\etal}(1993)]{kpi}
Bahcall, J. N., {\etal}
1993, \apjs, 87, 1

\bibitem[Bahcall {\etal}(1996)]{kpvii}
Bahcall, J. N., {\etal} 1996, \apj, 457, 19

\bibitem[Barlow \& Sargent(1997)]{bs97}
Barlow, T.  A. \& Sargent, W. L. W. 1997, \aj, 113, 136

\bibitem[Bergeron {\etal}(1983)]{bbdt83}
Bergeron, J., Boksenberg, A., Dennefeld, M., Tarenghi, M.
1983, \mnras, 202, 125

\bibitem[Crenshaw {\etal}(2000)]{cren00}
Crenshaw, D. M., Kraemer, S. B., Hutchings, J. B., Danks, A. C., Gull, T. R.,
Kaiser, M. E., Nelson, C. H., \& Weistrop, D.
2000, \apjl, 545, 27

\bibitem[Foltz {\etal}(1986)]{foltz86}
Foltz, C. B., Weymann, R. J., Peterson, B. P., Sun, L., Malkan, M. A.,
\& Chaffee, F. H. 1986,
\apj, 307, 504

\bibitem[Ganguly {\etal}(1999)]{gan99}
Ganguly, R., Eracleous, M., Charlton, J. C., \& Churchill, C. W. 1999,
\aj, 117, 2594

\bibitem[Ganguly {\etal}(2001)]{gan01}
Ganguly, R., Bond, N. A., Charlton, J. C., Eracleous, M.,  Brandt, W. N.,
Churchill, C. W.
2001, \apj, 549, 133

\bibitem[Halpern(1984)]{hal84}
Halpern, J. P.
1984, \apj, 281, 90

\bibitem[Hamann {\etal}(1995)]{ham95}
Hamann, F., Barlow, T. A., Beaver, E. A., Burbidge, E. M., Cohen, R. D,
Junkkarinen, V., \& Lyons, R.
1995, \apj, 443, 606

\bibitem[Hamann {\etal}(1997a)]{ham97a}
Hamann, F., Barlow, T. A., Junkkarinen, V., \& Burbidge, E. M. 1997a,
\apj, 478, 80

\bibitem[Hamann {\etal}(1997b)]{ham97b}
Hamann, F., Barlow, T. A., \& Junkkarinen, V. 1997b, \apj, 478, 87

\bibitem[Hamann {\etal}(1997c)]{ham97c}
Hamann, F., Barlow, T. A., Cohen, R. D., Junkkarinen, V., \& Burbidge,
E. M., 1997c, in ASP Conference Ser. 128, Mass Ejection from Active
Galactic Nuclei, ed. N. Arav, I. Shlosman, \& R. Weymann (San
Francisco: ASP), 19

\bibitem[Jannuzi {\etal}(1998)]{kpxiii}
Jannuzi, B. T., {\etal} 1998, \apjs, 118, 1

\bibitem[Maran {\etal}(1996)]{mar96}
Maran, S. P., Crenshaw, D. M., Mushotzky, R. F., Reichert, G. A.,
Carpenter, K. G., Smith, A. M., Hutchings, J. B., \& Weymann, R. J.
1996, \apj,465, 733

\bibitem[Monier {\etal}(2001)]{mon01}
Monier, E. M., Mathur, S., Wilkes, B., Elvis, M.
2001, astro-ph/0102348

\bibitem[Richards {\etal}(1999)]{rich99}
Richards, G. T., York, D. G., Yanny, B., Kollgaard, R. I.,
Laurent-Muehleisen, S. A., \& vanden Berk, D. E.
1999, \apj, 513, 576

\bibitem[Schneider {\etal}(1993)]{kpii}
Schneider, D. P., {\etal} 1993, \apjs, 87, 45

\bibitem[Shull \& Sachs(1993)]{ss93}
Shull, J. M., \& Sachs, E. R.
1993, \apj, 416, 536

\bibitem[Walter {\etal}(1990)]{wal90}
Walter, R., Courvoisier, T. J.-L., Ulrich, M.-H., \& Buson, L. M.
1990, \aap, 233, 53

\bibitem[Weymann {\etal}(1997)]{wsgh97}
Weymann, R. J., Morris, S. L., Gray, M. E., \& Hutchings, J. B.
1997, \apj, 483, 717

\end{thebibliography}
\end{document}